\documentclass[10pt]{article}
\usepackage{amsmath}
\usepackage{graphicx}
\usepackage{color}
\usepackage{orcidlink}
\usepackage{hyperref}
\hypersetup{colorlinks=true, linkcolor=blue, citecolor=blue, urlcolor=blue}
\usepackage{caption}
\usepackage{subcaption}
\begin{document}

\begin{center}
\textbf{\large{} Relativistic spin-0 Duffin-Kemmer-Petiau equation in Bonnor-Melvin-Lambda solution } 
\par\end{center}

\vspace{0.3cm}
 
\begin{center}
   
    \textbf{Faizuddin Ahmed\orcidlink{0000-0003-2196-9622}}\footnote{\textbf{faizuddinahmed15@gmail.com}}\\
    \vspace{0.10cm}
    \textit{Department of Physics, University of Science \& Technology Meghalaya, Ri-Bhoi, 793101, India}\\
     \vspace{0.3cm}
     \textbf{Abdelmalek Bouzenada\orcidlink{0000-0002-3363-980X}}\footnote{\textbf{abdelmalekbouzenada@gmail.com ; abdelmalek.bouzenada@univ-tebessa.dz}}\\
    \vspace{0.1cm}
    \textit{Laboratory of theoretical and applied Physics, Echahid Cheikh Larbi Tebessi University, Algeria} 
\par\end{center}

\vspace{0.3cm}

\begin{abstract}
In this paper, we conduct a comprehensive exploration of the relativistic quantum dynamics of spin-0 scalar particles, as described by the Duffin-Kemmer-Petiau (DKP) equation, within the framework of a magnetic space-time. Our focus is on the Bonnor-Melvin-Lambda (BML) solution, a four-dimensional magnetic universe characterized by a magnetic field that varies with axial distance. To initiate this investigation, we derive the radial equation using a suitable wave function ansatz and subsequently employ special functions to solve it. Furthermore, we extend our analysis to include Duffin-Kemmer-Petiau oscillator fields within the same BML space-time background. We derive the corresponding radial equation and solve it using special functions. Significantly, our results show that the geometry's topology and the cosmological constant (both are related with the magnetic field strength) influences the eigenvalue solution of spin-0 DKP fields and DKP-oscillator fields, leading to substantial modifications in the overall outcomes.
\end{abstract}

\vspace{0.1cm}

\textbf{keywords:} Exact solutions: Bonnor-Melvin space-time; Relativistic wave-equations; special functions; solutions of wave equations: bound states; 

\vspace{0.1cm}

\textbf{PACS numbers:} 04.20.Jb; 03.65.Pm ; 03.65.Ge ; 61.72.Lk ; 02.30.Gp; 03.65.Vf

\section{Introduction}

In the cosmic interplay of forces, the unyielding grasp of gravity on the grand scale collides with the capricious movements of particles on the quantum stage. It's a moment where the known laws of physics blur, unveiling a rich tapestry woven with threads of mystery and complexity. As we delve into this convergence of General Relativity (GR) and Quantum Mechanics (QM), we embark on a journey to uncover secrets that have eluded us for centuries. From the inception of black holes to the very fabric of spacetime itself, this encounter not only holds the promise of expanding our knowledge but also of pushing the boundaries of our comprehension. It beckons us to explore the essence of existence itself, where the distinct realms of gravity and quantum mechanics coalesce in a harmonious symphony of cosmic proportions. The interaction between gravitational forces and the dynamics of quantum mechanical systems is a captivating intellectual pursuit. Albert Einstein's groundbreaking general theory of relativity (GR) adeptly conceptualizes gravity as an intrinsic geometric feature of space-time \cite{k1}. This theory unveils a fascinating connection between space-time curvature and the emergence of classical gravitational fields, offering precise predictions for phenomena such as gravitational waves \cite{k2} and black holes \cite{k3}. Concurrently, the robust framework of quantum mechanics (QM) \cite{k4} provides invaluable insights into the nuanced behaviors of particles at the microscopic scale. The convergence of these two foundational theories compels us to delve into the profound mysteries that unfold at the intersection of the macroscopic realm governed by gravity and the quantum intricacies of the subatomic world. This interdisciplinary pursuit promises to unravel deeper layers of understanding about the fundamental nature of our universe, where the gravitational and quantum realms intricately dance in a cosmic ballet of profound significance.

The investigation into the impact of curved space on quantum mechanical problems, especially concerning gravitational field effects, stands as a central focus of research interest. The exploration of the interplay between gravity and quantum systems is an actively pursued area of study. In an effort to comprehend these interactions involving quantum particles, whether they be bosons or fermions, researchers have undertaken the challenge of solving both relativistic and non-relativistic wave equations. A myriad of methods has been employed by numerous authors to address these wave equations, resulting in a wealth of exact and approximate eigenvalue solutions. External magnetic and quantum flux fields have been introduced through minimal substitution in these equations, leading to subsequent solutions. Additionally, researchers have incorporated various potential models of physical significance into quantum systems, thereby enriching the results of solved wave equations. The exact and approximate eigenvalue solutions obtained serve as invaluable tools for refining models and numerical methods, especially when tackling intricate physical problems. The concept of the total wave function of quantum particles, encapsulating the physical properties of the quantum system, is well-established. These wave equations have been extensively explored within curved space-time geometries, including but not limited to the G\"{o}del Cosmological solutions \cite{key-1,key-2}, the Som-Raychaudhuri space-time \cite{key-2,key-3}, Schwarzschild-like solutions \cite{key-4}, and both topologically trivial \cite{key-5} and non-trivial \cite{key-6} space-time backgrounds. This broad exploration contributes to a deeper understanding of the intricate relationship between quantum mechanics and gravitational forces within diverse space-time contexts.

Numerous researchers have aimed into the exploration of relativistic wave equations within curved space-time, particularly in the presence of topological defects. These investigations encompass the study of wave equations involving cosmic strings, point-like global monopoles, cosmic strings with dislocation, screw dislocation as topological defects. The Klein-Gordon equation governs the dynamics of spin-0 scalar particles, while spin-1/2 particles are described by the Dirac equation. These wave equations have been scrutinized in diverse scenarios, including the hydrogen atom in cosmic string and point-like global monopole contexts \cite{key-7}, the spin-0 KG-oscillator field subjected to a magnetic field in cosmic string space-time \cite{key-8}, and the quantum motions of scalar and spin-half particles under a magnetic field in cosmic string space-time \cite{key-9}. Furthermore, non-relativistic wave equations have been explored in the context of topological defects, incorporating physical potential models (see Refs. \cite{key-10,key-11,key-12}). Another significant relativistic wave equation, akin to the Dirac equation, is the first-order Duffin-Kemmer-Petiau (DKP) equation. This equation accommodates both spin-0 and spin-1 fields or particles \cite{key-13,key-14,key-15,key-16}, proving useful in analyzing the relativistic interactions of spin-zero and spin-one hadrons with nuclei \cite{key-17}. Essentially, it serves as a direct extension of the Dirac equation, grounded in the DKP algebra \cite{key-18}. Numerous studies have explored applications of the DKP equation \cite{kk1,kk2,kk3,kk4,kk5,kk6,MH5,LBC,MH2,MH3,key-19, PRC, CPC, EPJP, IJTP, CJP, FBS, EPL2}.

In the domain of mathematical physics, the 1930s marked a significant milestone with the introduction of the Duffin–Kemmer–Petiau (DKP) algebra by R. J. Duffin, N. Kemmer, and G. Petiau \cite{key-13,key-14,key-15}. This algebra, later generalized by the DKP matrices, plays a crucial role in furnishing a relativistic framework for describing spin-0 and spin-1 particles via the DKP equation. Despite initially being categorized as a Dirac-like first-order equation, the DKP equation possesses an intricate algebraic structure. Until the 1970s, it was commonly assumed that the DKP equations were interchangeable with the Klein-Gordon (KG) equations and the Proca equations \cite{cor1}, yielding similar outcomes. As a result, interest in further developing the DKP equations was limited within the scientific community. However, subsequent revelations unveiled that this equivalence only holds under specific symmetrical conditions \cite{cor2}. Further investigation revealed that in scenarios where this symmetry was broken, solutions derived from the DKP equation diverged from those of the other two equations. These discoveries reignited scholarly curiosity in the DKP equations, sparking a resurgence of interest in the topic.

The exploration of magnetic fields within the framework of general relativity has been a subject of study, opening new avenues of inquiry. Solutions to the Einstein-Maxwell equations grace the scientific stage, including the elegant Manko solution \cite{k18,k19}, the Bonnor-Melvin universe \cite{k20,k21}, and a recent proposition \cite{k22} introducing the cosmological constant into the Bonnor-Melvin narrative. Shifting our focus to the intersection of general relativity and quantum physics, a pivotal consideration is how these two theories connect. The study of the Klein-Gordon and Dirac equations within the curvature effects produced by space-time has been explored \cite{k23}. This intricate connection extends to diverse scenarios, from particles pirouetting in the gravitational embrace of Schwarzschild \cite{k25} and Kerr black holes \cite{k26} to the quantum harmonies resonating in cosmic string backdrops \cite{k27,k28,k29}. It envelops quantum oscillators \cite{k30, k31, key-6, k35, k35-1, k35-2, k35-3}, contemplates the ethereal echoes of the Casimir effect \cite{k36,k37}, and observes the rhythmic interplay of particles within the cosmic symphony of the Hartle-Thorne space-time \cite{k38}. These ventures, among others \cite{k39,k40,k41}, have birthed profound insights into the intricate ballet of quantum systems pirouetting through the arbitrary geometries of space-time. Consequently, a tantalizing avenue unfolds with the examination of quantum particles pirouetting within a space-time sculpted by the guiding strokes of a magnetic field. A poetic precursor in \cite{k42} explored Dirac particles in the magnetic embrace of the Melvin metric. Meanwhile, our current endeavor embarks on a symphony of inquiry, unraveling the enigma of spin-0 bosons pirouetting within the magnetic universe, gracefully incorporating the cosmological constant, as suggested in \cite{k22}.

The Klein-Gordon equation (KG) and the Duffin-Kemmer-Petiau (DKP) equation diverge in their treatment of spin and algebraic structure. While the KG equation is tailored for spinless particles, the DKP equation accommodates both spin-0 and spin-1 particles through its intricate algebraic framework involving DKP matrices. Despite their shared adherence to relativistic principles, the DKP equation offers a more encompassing description of particles endowed with intrinsic angular momentum. Initially believed to be equivalent under specific symmetrical conditions, further scrutiny has unveiled distinct solutions arising from the DKP equation when symmetry is broken. This underscores the subtle discrepancies between the two equations. Despite historical favor towards the KG equation, recent advancements have reignited interest in the DKP equation, prompting its exploration across various research domains within the scientific community. The relativistic DKP equation for charged free scalar bosons with mass M in curved space, as a first-order formulation, is provided by the following wave equation \cite{key-13,key-14,key-15,key-16}.
\begin{equation}
\left(i\,\widetilde{\beta}^{\mu}\,\nabla_{\mu}-M\right)\Psi=0,\label{eq:1}
\end{equation}
where $\widetilde{\beta}^{\mu}=e^{\mu}_{a}\,\beta^a$ are the DKP-matrices in curved space and $\beta^a$ is the DKP matrices in flat space. The matrices $\widetilde{\beta}^{\mu}$ satisfy the following commutation rules
\begin{eqnarray}
\widetilde{\beta}^{\kappa}\,\widetilde{\beta}^{\nu}\,\widetilde{\beta}^{\lambda}+\widetilde{\beta}^{\lambda}\,\widetilde{\beta}^{\nu}\,\widetilde{\beta}^{\kappa}=g^{\kappa\nu}\,\widetilde{\beta}^{\lambda}+g^{\nu\lambda}\,\widetilde{\beta}^{\kappa}.\label{eq:2}
\end{eqnarray}
In equation (\ref{eq:1}) $\nabla_{\mu}$ is the covariant derivative defined by
\begin{equation}
\nabla_{\mu}=\partial_{\mu}+\Gamma_{\mu}\label{eq:3}
\end{equation}
with $\Gamma_{\mu}$ is the affine connection related with spin connection $\omega_{\mu\, ab}$ defined by
\begin{equation}
\Gamma_{\mu}=\frac{1}{2}\,\omega_{\mu\, ab}\,\left[\beta^{a},\beta^{b}\right].\label{eq:4}
\end{equation}
Here this spin connection in terms of the Christoffel symbols of second kind is defined by
\begin{equation}
\omega_{\mu\,\,b}^{a}=e_{\tau}^{a}\,e_{b}^{\nu}\,\Gamma_{\mu\nu}^{\tau}-e_{b}^{\nu}\,\partial_{\mu}\,e_{\tau}^{a},\label{eq:5}
\end{equation}
where the Christoffel symbols of second kind is given by
\begin{equation}
\Gamma_{\mu\nu}^{\tau}=\frac{1}{2}\,g^{\tau\kappa}\,\left(g_{\nu\kappa,\mu}+g_{\mu\kappa,\nu}- g_{\mu\nu,\kappa}\right).\label{eq:6}
\end{equation}
The DKP-matrices in flat space are given by the following $5 \times 5$ matrices as follows \cite{kk1,MH3}:
\begin{eqnarray}
    &&\beta^0=\left(\begin{array}{cc}
         \kappa & {\bf 0}_{2 \times 3} \\
         {\bf 0}_{3 \times 2} & {\bf 0}_{3\times 3} 
    \end{array}
    \right),\quad
    \beta^i=\left(\begin{array}{cc}
         {\bf 0}_{2 \times 2} & \vec{\gamma}\\
         -\vec{\gamma}^{T} & {\bf 0}_{3 \times 3} 
    \end{array}
    \right),\quad \kappa=\left(\begin{array}{cc}
        0 & 1 \\
        1 &  0
    \end{array}
    \right),\nonumber\\
    &&\gamma^1=\left(\begin{array}{ccc}
        -1 & 0 & 0 \\
         0 & 0 & 0
    \end{array}
    \right),\quad 
    \gamma^2=\left(\begin{array}{ccc}
        0 & -1 & 0 \\
         0 & 0 & 0
    \end{array}
    \right),\quad 
    \gamma^1=\left(\begin{array}{ccc}
        0 & 0 & -1 \\
         0 & 0 & 0
    \end{array}
    \right).\label{1}
\end{eqnarray}

Our objective is to explore the relativistic quantum dynamics of spin-0 scalar bosons described by the DKP equation within the framework of a magnetic universe. We specifically focused on a four-dimensional magnetic space-time with a cosmological constant called Bonnor-Melvin-Lambda (BML) solution. Afterwards, we investigate the DKP-oscillator in the same BML space-time background. In both cases, we solve the wave equation and obtain the relativistic energy spectrum of spin-0 DKP field and DKP-oscillator fields. In fact, we see that the topology of the geometry and cosmological constant influences the energy profiles of the quantum systems and gets modification compared to Landau levels.

The structure of this paper is as follows: In {\it section 2}, we derive the necessary physical quantities involved in the DKP-equation in BML space-time background. We then solve this radial equation through special functions and presents the energy profile of spin-0 DKP fields. In {\it section 3}, we focus on DKP-oscillator fields in the background of same BML space-time. We derive and solve the radial equation through special functions. In {\it section 4}, we present our results and discussion. Throughout the paper, we choose the system of units, where $c=1=\hbar$.

\section{DKP-Equation in Bonnor-Melvin-Lambda solution}

In this section, our attention shifts towards the quantum motions of spin-0 Bosons described by the DKP equation within the context of the Bonnor-Melvin-Lambda space-time, enriched with the presence of a cosmological constant. The metric in focus represents a static solution with cylindrical symmetry, derived from Einstein's equations. This metric is intricately influenced by a homogeneous magnetic field, presenting an intriguing backdrop for our exploration. The magnetic space-time in cylindrical system is described by the following line-element \cite{k22,MA2}.
\begin{equation}
ds^{2}=g_{\mu\nu}\,dx^{\mu}\,dx^{\nu}=-dt^{2}+d\rho^{2}+\sigma^{2}\,\sin^{2}\left(\sqrt{2\Lambda}\,\rho\right)\,d\varphi^{2}+dz^{2},\label{eq:7}
\end{equation}
where $\sigma>0$ is the topology parameter which produces an angular deficit and $\Lambda$ is the positive cosmological constant. The magnetic field strength associated with this metric is given by $H=|\sigma|\,\sqrt{\Lambda}\,\sin \left(\sqrt{2\Lambda}\,\rho\right)$ which acts along the $z$-direction.

The covariant metric tensor $g_{\mu\nu}$ and it's contravariant form $g^{\mu\nu}$ for the line-element (\ref{eq:7}) are given by
\begin{equation}
g_{\mu\nu}=\left(\begin{array}{cccc}
-1 & 0 & 0 & 0\\
0 & 1 & 0 & 0\\
0 & 0 & \sigma^2\,\sin^2\left(\sqrt{2\Lambda}\,\rho\right) & 0\\
0 & 0 & 0 & 1
\end{array}\right),\quad 
g^{\mu\nu}=\left(\begin{array}{cccc}
-1 & 0 & 0 & 0\\
0 & 1 & 0 & 0\\
0 & 0 & \frac{1}{\sigma^2\,\sin^2\left(\sqrt{2\Lambda}\,\rho\right)} & 0\\
0 & 0 & 0 & 1
\end{array}\right)\,.\label{eq:8}
\end{equation}

Now, we construct the tetrads $e^{\mu}_{a}$ and its covariant form for the metric tensor (\ref{eq:8}) are given by 
\begin{equation}
e_{\mu}^{a}\left(\boldsymbol{x}\right)=\left(\begin{array}{cccc}
1 & 0 & 0 & 0\\
0 & 1 & 0 & 0\\
0 & 0 & |\sigma|\,\sin\left(\sqrt{2\Lambda}\rho\right) & 0\\
0 & 0 & 0 & 1
\end{array}\right),\,\,\,\,\,\,e_{a}^{\mu}\left(\boldsymbol{x}\right)=\left(\begin{array}{cccc}
1 & 0 & 0 & 0\\
0 & 1 & 0 & 0\\
0 & 0 & \frac{1}{|\sigma|\,\sin\left(\sqrt{2\Lambda}\rho\right)} & 0\\
0 & 0 & 0 & 1
\end{array}\right)\label{eq:10}
\end{equation}
such that $e^{\mu}_{a}\,e_{\nu}^{a}=\delta^{\mu}_{\nu}$ and $e_{\mu}^{a}\,e^{\mu}_{b}=\delta^{a}_{b}$.

The non-zero comonents of the Christoffel symbols of the second kind using the metric tensor (\ref{eq:8}) are given by  
\begin{eqnarray}
&&\Gamma_{\mu\nu}^{\rho}=\left(\begin{array}{cccc}
0 & 0 & 0 & 0\\
0 & 0 & 0 & 0\\
0 & 0 & -\sigma^{2}\,\sqrt{\frac{\Lambda}{2}}\,\sin\left(2\,\sqrt{2\Lambda}\,\rho\right) & 0\\
0 & 0 & 0 & 0
\end{array}\right),\nonumber\\
&&\Gamma_{\mu\nu}^{\varphi}=\left(\begin{array}{cccc}
0 & 0 & 0 & 0\\
0 & 0 & \frac{\sqrt{2\Lambda}}{\tan\left(\sqrt{2\Lambda}\rho\right)} & 0\\
0 & \frac{\sqrt{2\Lambda}}{\tan\left(\sqrt{2\Lambda}\rho\right)} & 0 & 0\\
0 & 0 & 0 & 0
\end{array}\right).\label{eq:11}
\end{eqnarray}
The non-zero components of spin connections are given by
\begin{equation}
\omega_{\varphi ab}=\left(\begin{array}{cccc}
0 & 0 & 0 & 0\\
0 & 0 & |\sigma|\sqrt{2\Lambda}\cos\left(\sqrt{2\Lambda}\rho\right) & 0\\
0 & -|\sigma|\sqrt{2\Lambda}\cos\left(\sqrt{2\Lambda}\rho\right) & 0 & 0\\
0 & 0 & 0 & 0
\end{array}\right).\label{eq:12}
\end{equation}

Let, the total wave function is of the following form
\begin{equation}
\Psi_{DKP} (t, \rho, \phi, z)=e^{-i\,E\,t+i\,k\,z+i\,\ell\,\varphi}\,\left(\psi_{1}\,\,\psi_{2}\,\,\psi_{3}\,\,\psi_{4}\,\,\psi_{5}\right)^{T},\label{eq:14}
\end{equation}
where $E$ is the particles energy, $\ell$ is the angular quantum number, $k$ is an arbitrary constant, and $\psi_i=\psi_i (\rho)$ is the radial wave function with the five-component DKP spinor $\left(\psi_{1}\,\,\psi_{2}\,\,\psi_{3}\,\,\psi_{4}\,\,\psi_{5}\right)^{T}$.

Therefore, the DKP-equation (\ref{eq:1}) in BML space-time can be explicitly written as
\begin{eqnarray}
\Bigg[i\beta^{0}\partial_{t}+i\beta^{1}\partial_{\rho}+i\frac{\beta^{2}}{|\sigma|\sin\left(\sqrt{2\Lambda}\rho\right)}\left\{\partial_{\varphi}-|\sigma|\left(\sqrt{2\Lambda}\right)\cos\left(\sqrt{2\Lambda}\rho\right)\left(\beta^{1}\beta^{2}-\beta^{2}\beta^{1}\right)\right\}+i\beta^{3}\partial_{z}-M\Bigg]\psi=0,\label{eq:17}
\end{eqnarray}

Substituting the wave function (\ref{eq:14}) and the DKP flat matrices $\beta^a$ into the equation (\ref{eq:17}) results the following set of equations: 
\begin{eqnarray}
&&E\,\psi_{2}-i\,\left(\partial_{1}+\frac{\left(\sqrt{2\Lambda}\right)}{\tan\left(\sqrt{2\Lambda}\rho\right)}\right)\,\psi_{3}+\left(\frac{\ell}{|\sigma|\, \sin\left(\sqrt{2\Lambda}\rho\right)}\right)\,\psi_{4}+k\,\psi_{5}=M\,\psi_{1}.\label{eq:18}\\
&&E\,\psi_{1}=M\,\psi_{2}.\label{eq:19}\\
&&i\,\partial_{1}\,\psi_{1}=M\,\psi_{3}.\label{eq:20}\\
&&\frac{-\ell}{|\sigma|\,\sin\left(\sqrt{2\Lambda}\rho\right)}\,\psi_{1}=M\,\psi_{4}.\label{eq:21}\\
&&-k\,\psi_{1}=M\,\psi_{5}.\label{eq:22}
\end{eqnarray}

Decoupled the above set of equations, we obtain the following differential equation in terms $\psi_1$ :
\begin{equation}
    \left[\frac{\partial^{2}}{\partial\rho^{2}}+\frac{\left(\sqrt{2\Lambda}\right)}{\tan\left(\sqrt{2\Lambda}\rho\right)}\frac{\partial}{\partial\rho}+E^{2}-M^{2}-k^{2}-\frac{\ell^{2}}{\sigma^{2}\,\sin^{2}\left(\sqrt{2\Lambda}\rho\right)}\right]\psi_{1}=0.\label{eq:23}
\end{equation}
To solve the above differential equation, we employ approximation to the trigonometric functions upto first order, we obtain
\begin{equation}
    \left[\frac{\partial^{2}}{\partial\rho^{2}}+\frac{1}{\rho}\,\frac{\partial}{\partial\rho}+E^{2}-M^{2}-k^{2}-\frac{\tau^2}{\rho^2}\right]\,\psi_{1}=0.\label{eq:24}
\end{equation}
where
\begin{equation}
    \tau=\frac{|\ell|}{\sqrt{2\,\Lambda}\,|\sigma|}.\label{eq:25}
\end{equation}
Equation (\ref{eq:23}) can be expressed into a standard differential equation form as follows:
\begin{equation}
    \rho^2\,\psi''_{1}+\rho\,\psi'_{1}+(\lambda^2\,\rho^2-\tau^2)\,\psi_1=0,\label{eq:26}
\end{equation}
where $\lambda=\sqrt{E^2-M^2-k^2}$.

\begin{center}
\begin{figure}
\begin{centering}
\subfloat[$|\sigma|=0.1, \ell=1$]{\centering{}\includegraphics[scale=0.5]{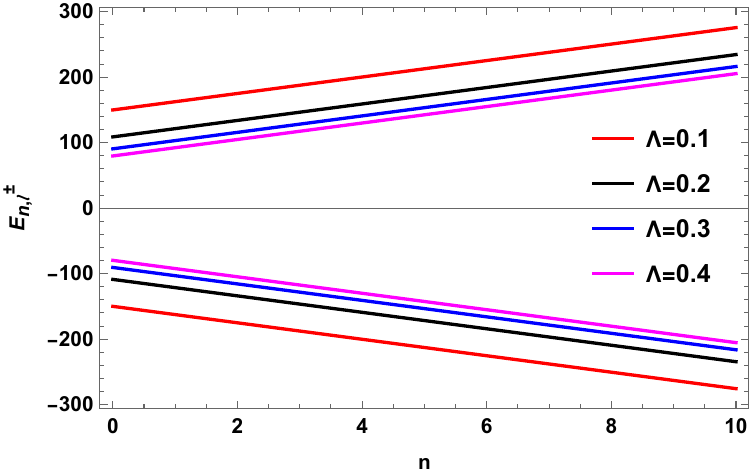}}\quad\quad
\subfloat[$\Lambda=0.1, \ell=1$]{\centering{}\includegraphics[scale=0.5]{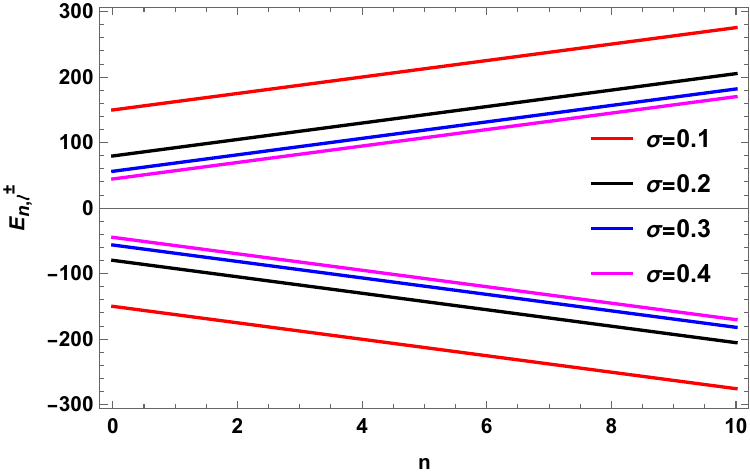}}
\par\end{centering}
\begin{centering}
\subfloat[$|\sigma|=0.5, \Lambda=0.2$]{\centering{}\includegraphics[scale=0.5]{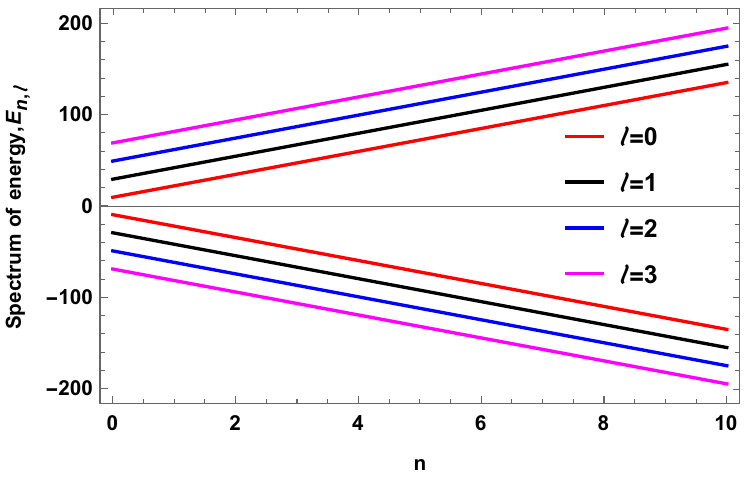}}
\par\end{centering}
\caption{The energy spectrum $E_{n,\ell}^{\pm}$ for the relation given in
equation (\ref{eq:29}), where the parameters are
set as $k=M=1$ and $\rho_{0}=0.25$.}
\begin{centering}
\subfloat[$|\sigma|=0.2, \ell=1$]{\centering{}\includegraphics[scale=0.5]{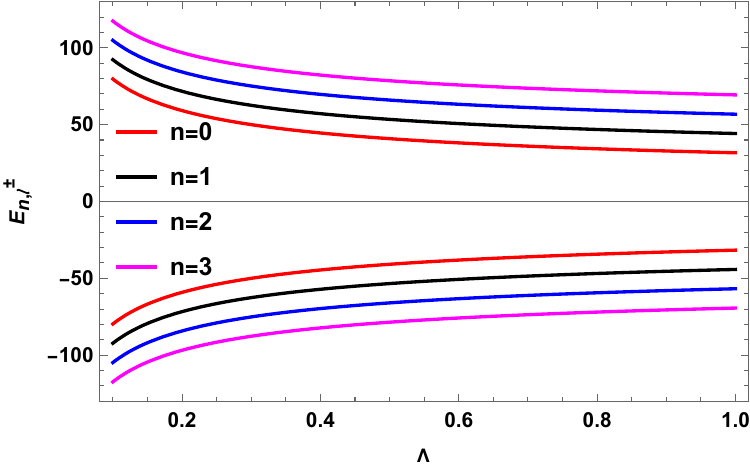}}\quad\quad
\subfloat[$\Lambda=0.2, \ell=1$]{\centering{}\includegraphics[scale=0.5]{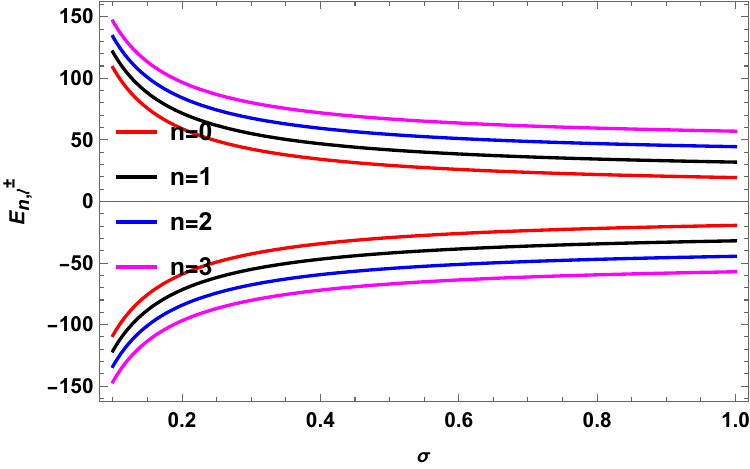}}
\par\end{centering}
\caption{The energy spectrum $E_{n,\ell}^{\pm}$ for the relation given in
equation (\ref{eq:29}), where the parameters are
set as $k=M=1$ and $\rho_{0}=0.25$.}
\end{figure}
\par\end{center}

Equation (\ref{eq:26}) is the Bessel second-order differential equation whose solution is well-known. The solution is $\psi_1=c_1\,J_{\tau}(\Lambda\,\rho)+c_2\,Y_{\tau}(\lambda\,\rho)$ \cite{MA,GBA}, where $J_{\tau}$ and $Y_{\tau}$ are the Bessel functions of the first and second kind, respectively. It is also well-known that the Bessel function of the second kind is undefined at the origin $\rho=0$ but first kind is finite. Therefore, the requirement of the wave function $\psi_1(\rho)$ at the origin implies that the constant $c_2 \to 0$. Therefore, the regular solution of the above differential equation is given by
\begin{equation}
    \psi_1 (\rho)=c_1\,J_{\tau}(\lambda\,\rho)\,.\label{eq:27}
\end{equation}

In this scenario, we consider the motion of a spin-0 bosonic field confined within a finite region of space, where a hard-wall confining potential is present. This confinement is particularly significant as it offers an excellent approximation for investigating the quantum properties of systems, such as gas molecules and other particles inherently constrained within a defined spatial domain. The hard-wall confinement is characterized by a condition stipulating that at a certain axial distance, denoted by $\rho=\rho_0$, the radial wave function $\Psi_1$ becomes zero, i.e., $\Psi_1 (\rho=\rho_0)=0$.

The study of the hard-wall potential has proven valuable in various contexts, including investigations of the Klein-Gordon oscillator in cosmic string space-time \cite{k28}, scalar fields under non-inertial effects \cite{k29}, the Dirac neutral particle \cite{KB, KB2, KB3}, the harmonic oscillator in an elastic medium \cite{KB5}, systems analogous to the Landau-Aharonov-Casher system in cosmic string space-time \cite{KB4}, and on a Dirac neutral particle in analogous way to a quantum dot \cite{KB6}, a neutral particle with no permanent electric dipole moment \cite{KB7}, and in a Landau-type quantization from a Lorentz symmetry violation background \cite{KB8}. Additionally, it has shed light on the behavior of Dirac and Klein-Gordon oscillators \cite{RLLV5, SR}. This exploration of the hard-wall confining potential in diverse scenarios enriches our understanding of its impact on quantum systems, providing insights into the behavior of scalar and oscillator fields subject to this form of confinement.

To obtain energy levels of the bound states from the boundary condition, we take a fixed $\rho=\rho_0$ that is sufficiently large such that we can consider $\lambda,\rho_0>>1$. Hence, by accepting this approximation $\lambda,\rho_0>>1$, we can express the Bessel function of the first kind in the form \cite{MA,GBA}:
\begin{equation}
    J_{\iota} (\lambda\,\rho_0) \propto \cos \Big(\lambda\,\rho_0-\frac{\tau\,\pi}{2}-\frac{\pi}{4}\Big).\label{eq:28}
\end{equation}

Therefore, substituting this asymptotic form (\ref{eq:28}) into the equation (\ref{eq:27}) and using the boundary condition $\psi_1 (\rho=\rho_0)=0$, after simplification, we obtain the following expression of the energy eigenvalue given by
\begin{equation}
    E_{n,\ell}=\pm\,\sqrt{M^2+k^2+\Big(2\,n+\frac{3}{2}+\frac{|\ell|}{\sqrt{2\,\Lambda}\,|\sigma|}\Big)^2\,\frac{\pi^2}{4\,\rho^2_{0}}\Bigg]},\label{eq:29}
\end{equation}
where $n=0, 1,2,3,....$.

Equation (\ref{eq:29}) describe the relativistic energy eigenvalue of spin-0 bosonic fields within the Bonnor-Melvin-Lambda space-time background in the presence of hard-wall confining potential. It's evident that the cosmological constant $\Lambda$ and the topological parameter $|\sigma|$ associated with the space-time geometry modify the energy eigenvalue. Additionally, the energy levels are influenced by the quantum numbers $\{n, \ell\}$. To elucidate the influence of these parameters on the energy levels, we've generated Figures 1-2.

In Figure 1,  we illustrate the energy eigenvalue (\ref{eq:29}) of spin-0 bosonic fields, varying the values of the cosmological constant $\Lambda$ in Fig. 1(a), the topology parameter $|\sigma|$ in Fig. 1(b), and the orbital quantum number $\ell$ in Fig. 1(c). In Figs. 1(a)-(b), a linear increase in the energy level's nature is visible, which shifts downward with increasing $\Lambda$ and $|\sigma|$. Conversely, in Fig. 1(c), this trend shifts upward with increasing $\ell$. In Figure 2, the energy spectrum concerning the cosmological constant $\Lambda$ in Fig. 2(a), and the topology parameter $|\sigma|$ in Fig. 2(b) for fixed quantum number $n$, gradually decreases. This decreasing trend shifts upward with increasing values of the quantum number $n$.

\section{DKP-oscillator in Bonnor-Melvin-Lambda solution}

In this section, we study the relativistic quantum dynamics of spin-0 oscillator field described by the Duffin-Kemmer-Petiau (DKP) oscillator fields in the context of Bonnor-Melvin-Lambda space-time backgrounds. This DKP-oscillator is studied by a nonminimal substitution via $-i\,\vec{\nabla}_{\alpha} \to \Big(-i\,\vec{\nabla}_{\alpha}-i\,M\,\omega\,\eta^0\,\vec{\rho}\Big)$ or the radial momentum operator $\partial_{\rho} \to (\partial_{\rho}+M\,\omega\,\eta^{0}\,\rho)$ into the DKP equation. Here $\omega$ is the oscillator frequency and $\eta^0$ is related with $\beta^0$ matrix. It is worth mentioning that this DKP-oscillator has been investigated by numerous authors in various curved space-times background, such as G\"{o}del and G\"{o}del-type solutions in absence of cosmic strings \cite{CTP}, topological defects background produced by cosmic strings and global monopoles \cite{kk1, LBC, MH2, MH3, MH, AB2, NC}, in the presence of minimal length  \cite{JMP2,JMP3}, in a non-commutative phase space background \cite{EPJC, GDM, MF}, and in other scenario \cite{PHYSSR, PLA}. 

In terms of four-vector, this nonminimal substitution will be
\begin{equation}
 \partial_{\mu}\rightarrow\partial_{\mu}+M\omega X_{\mu}\eta^{0}, \label{osci:1}
\end{equation}
where the four-vector is defined by
\begin{equation}
X_{\mu}=\left(0,\rho,0,0\right). \label{osci:2}
\end{equation}
And
\begin{equation}
\eta^{0}=2\left(\beta^{0}\right)^{2}-{\bf I}_{5\times 5}. \label{osci:3}
\end{equation}

Therefore, the DKP-oscillator equation is described by 
\begin{equation}
 \left[i\,\widetilde{\beta}^{\mu}\left(\partial_{\mu}+M\,\omega\, X_{\mu}\,\eta^{0}+\frac{1}{2}\,\omega_{\mu ab}\,\left[\beta^{a},\beta^{b}\right]\,\right)-M\right]\psi=0. \label{osci:4}
\end{equation}

Explicitly writing this equation (\ref{osci:4}) in the space-time background (\ref{eq:7}), we obtain
\begin{equation}
\left[i\,\widetilde{\beta}^{0}\,\partial_{t}+i\,\widetilde{\beta}^{1}\,\left(\partial_{\rho}+M\,\omega\,\rho\,\eta^{0}\right)+i\,\widetilde{\beta}^{2}\left(\partial_{\varphi}+\frac{1}{2}\,\omega_{\varphi\,ab}\,\left[\beta^{a},\beta^{b}\right]\right)+i\,\widetilde{\beta}^{3}\,\partial_{z}-M\right]\psi=0. \label{osci:5}
\end{equation}
That can be written as
\begin{eqnarray}
&&\Bigg[i\,\beta^{0}\,\partial_{t}+i\,\beta^{1}\,(\partial_{\rho}+M\,\omega\,\eta^0\,\rho)+i\frac{\beta^{2}}{|\sigma|\,\sin\left(\sqrt{2\Lambda}\rho\right)}\left\{\partial_{\varphi}-|\sigma|\,\left(\sqrt{2\Lambda}\right)\,\cos\left(\sqrt{2\Lambda}\rho\right)\left(\beta^{1}\beta^{2}-\beta^{2}\beta^{1}\right)\right\}\nonumber\\
&&+i\beta^{3}\partial_{z}-M\Bigg]\psi=0,\label{osci:sp}
\end{eqnarray}

Substituting the wave function (\ref{eq:14}) and flat $\beta^a$ matrices, we obtain the following set of equations:
\begin{eqnarray}
&&E\,\psi_{2}-i\,\left[\partial_{\rho}+M\,\omega\,\rho+\frac{\left(\sqrt{2\Lambda}\right)}{\tan\left(\sqrt{2\Lambda}\rho\right)}\right]\,\psi_{3}+\left[\frac{\ell}{|\sigma|\, \sin\left(\sqrt{2\Lambda}\rho\right)}\right]\,\psi_{4}+k\,\psi_{5}=M\,\psi_{1}. \label{osci:6}\\
&&E\,\psi_{1}=M\,\psi_{2}. \label{osci:7}\\
&&i\,\left(\partial_{\rho}-M\,\omega\,\rho\right)\,\psi_{1}=M\,\psi_{3}.\label{osci:8}\\
&&-\frac{|\ell|}{|\sigma|\,\sin\left(\sqrt{2\Lambda}\rho\right)}\,\psi_{1}=M\,\psi_{4}. \label{osci:9}\\
&&-k\,\psi_{1}=M\,\psi_{5}. \label{osci:10}
\end{eqnarray}

Decoupled the above set of equations, we obtain the following differential equation:
\begin{equation}
\left[\left(\partial_{\rho}+M\,\omega\,\rho+\frac{\left(\sqrt{2\Lambda}\right)}{\tan\left(\sqrt{2\Lambda}\rho\right)}\right)\left(\partial_{\rho}-M\,\omega\,\rho\right)-\frac{\ell^{2}}{\sigma^{2}\,\sin^{2}\left(\sqrt{2\Lambda}\rho\right)}+\lambda^{2}\right]\psi_{1}=0. \label{osci:11}
\end{equation}
That can be written as
\begin{equation}
\left[\frac{\partial^{2}}{\partial\rho^{2}}+\frac{\left(\sqrt{2\Lambda}\right)}{\tan\left(\sqrt{2\Lambda}\rho\right)}\frac{\partial}{\partial\rho}-\left(M\,\omega\,\rho\right)^{2}-\frac{\ell^{2}}{\sigma^{2}\,\sin^{2}\left(\sqrt{2\Lambda}\rho\right)}-\frac{M\,\omega\,\rho\,\left(\sqrt{2\Lambda}\right)}{\tan\left(\sqrt{2\Lambda}\rho\right)}+\lambda^2-M\,\omega\right]\psi_{1}=0 \label{osci:12}
\end{equation}
where $\lambda=\sqrt{E^2-M^2-k^2}$.

To solve the above second-order differential equation, we employ approximation into the trigonometric functions up to the first order. However, one can try to find general solution of this differential equation by doing a substitution $u=\cos \left(\sqrt{2\Lambda}\rho\right)$ which we omitted here. Therefore, the above differential equation after approximation becomes
\begin{equation}
    \left[\frac{\partial^{2}}{\partial\rho^{2}}+\frac{1}{\rho}\,\frac{\partial}{\partial\rho}-M^2\,\omega^2\,\rho^2-\frac{\tau^2}{\rho^2}+\xi\right]\psi_{1}=0, \label{osci:13}
\end{equation}
where
\begin{equation}
    \tau=\frac{|\ell|/|\sigma|}{\sqrt{2\Lambda}},\quad \xi=\lambda^2-2\,M\,\omega.\label{osci:14}
\end{equation}

Transforming to a new variable via $s=M\,\omega\,\rho^2$ into the above equation (\ref{osci:13}) results the following differential equation form:
\begin{equation}
\psi''_{1}(s)+\frac{(d_1-\,d_2\,s)}{s\,(1-d_3\,s)}\,\psi'_{1} (s)+\frac{\Big(-\zeta_1\,s^2+\zeta_2\,s-\zeta_3\Big)}{s^2\,(1-d_3\,s)^2}\,\psi_{1} (s)=0,\label{osci:15}
\end{equation}
where $d_1=1$, $d_2=0=d_3$ and 
\begin{equation}
\zeta_1=\frac{1}{4},\quad \zeta_2=\frac{\xi}{4\,M\,\omega},\quad \zeta_3=\frac{\tau^2}{4}.\label{osci:16}
\end{equation}
Equation (\ref{osci:15}) is homogeneous second-order differential equation which can be solves using the Nikiforov-Uvarov method \cite{AFN}. Following this method, we obtain the following expression of the energy spectrum  given by
\begin{equation}
    E_{n,\ell}=\pm\,\sqrt{M^2+k^2+4\,M\,\omega\,\Bigg(n+1+\frac{|\ell|}{2\,\sqrt{2\Lambda}\,|\sigma|}\Bigg)}.\label{osci:17}
\end{equation}

\begin{center}
\begin{figure}
\begin{centering}
\subfloat[$|\sigma|=0.1,\ell=1=\omega$]{\centering{}\includegraphics[scale=0.5]{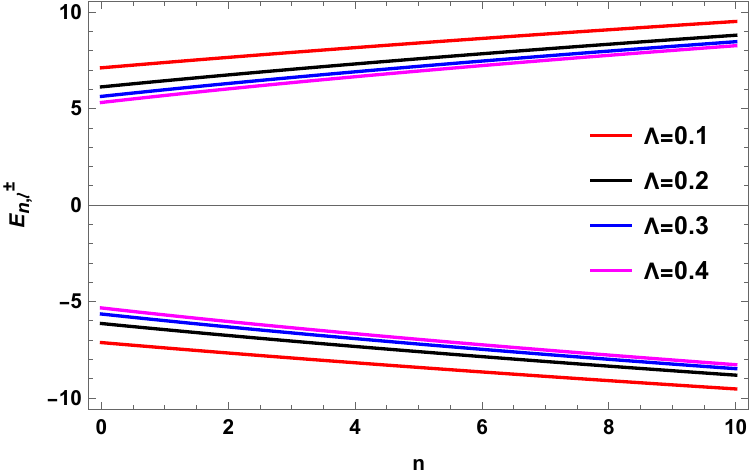}}\quad\quad
\subfloat[$\Lambda=0.1,\ell=1=\omega$]{\centering{}\includegraphics[scale=0.5]{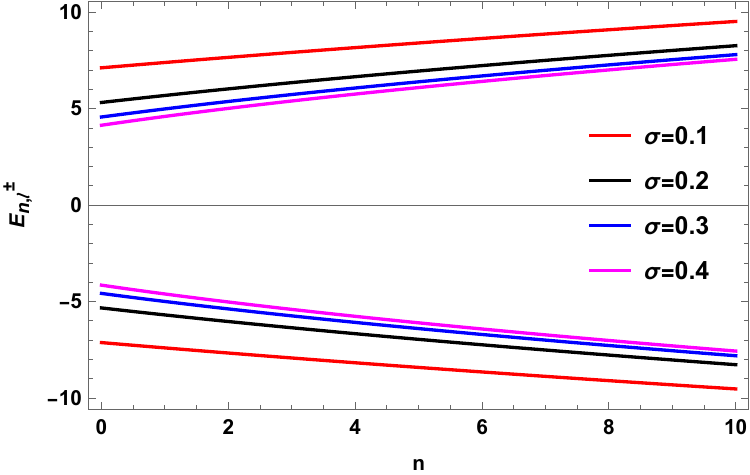}}
\par\end{centering}
\begin{centering}
\subfloat[$|\sigma|=0.5,\Lambda=0.2, \omega=1$]{\centering{}\includegraphics[scale=0.5]{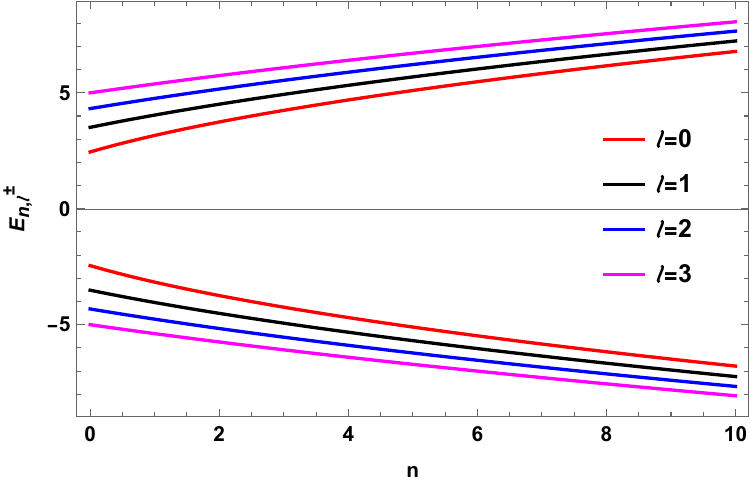}}\quad\quad
\subfloat[$|\sigma|=0.5,\Lambda=0.2,\ell=1$]{\centering{}\includegraphics[scale=0.5]{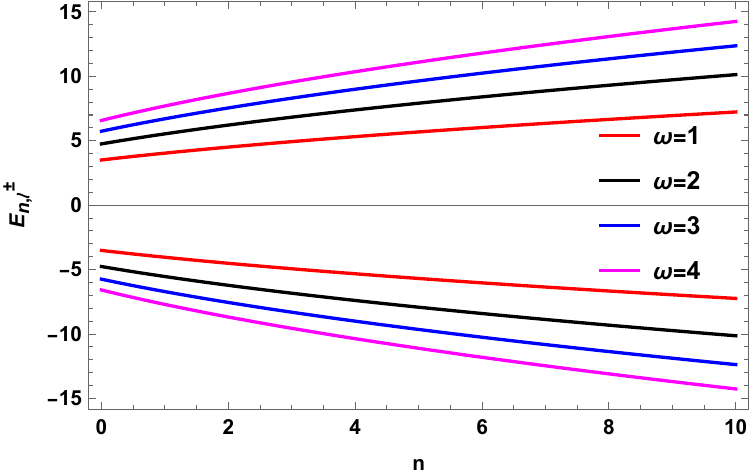}}
\par\end{centering}
\caption{The energy spectrum $E_{n,\ell}^{\pm}$ for the relation given in
equation (\ref{osci:17}), where the parameters are
set as $k=M=1$ .}
\begin{centering}
\subfloat[$|\sigma|=0.2,\ell=1=\omega$]{\centering{}\includegraphics[scale=0.5]{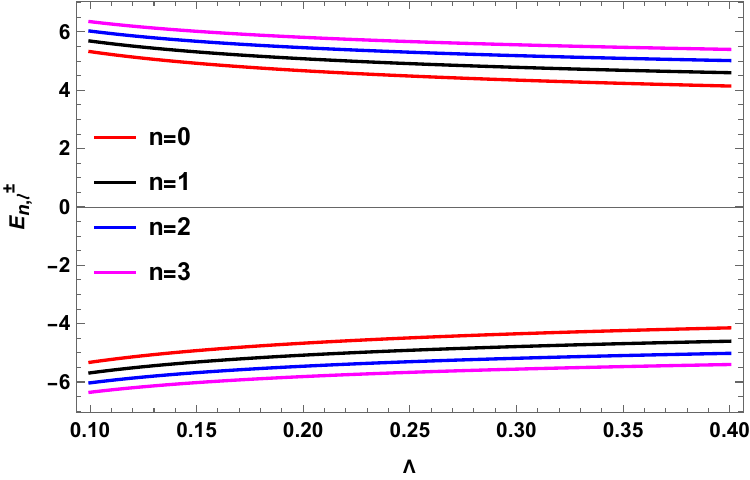}}\quad\quad
\subfloat[$\Lambda=0.2,\ell=1=\omega$]{\centering{}\includegraphics[scale=0.5]{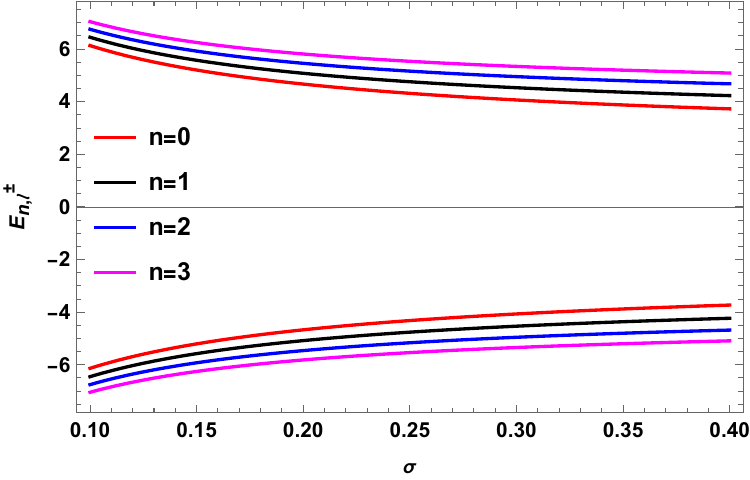}}
\par\end{centering}
\caption{The energy spectrum $E_{n,\ell}^{\pm}$ for the relation given in
equation (\ref{osci:17}), where the parameters are
set as $k=M=1$.}
\end{figure}
\par\end{center}

The corresponding radial wave function will be
\begin{equation}
    \psi_{1\,n,\ell}(\rho)=\mathcal{N}\,\left(M\,\omega\,\rho^2\right)^{\frac{|\ell|}{2\,\sqrt{2\Lambda}\,|\sigma|}}\,Exp\left[-\frac{1}{2}\,M\,\omega\,\rho^2\right]\,L^{\Big(\frac{|\ell|}{2\,\sqrt{2\Lambda}\,|\sigma|}\Big)}_{n} (M\,\omega\,\rho^2),\label{osci:18}
\end{equation}
where $\mathcal{N}$ is the normalization constant.

Therefore, other components of the wave functions are given by
\begin{eqnarray}
&&\psi_{2\,n,\ell}(\rho)=\frac{E_{n,\ell}}{M}\,\psi_{1\,n,\ell}.\label{osci:19}\\
&&\psi_{3\,n,\ell}(\rho)=\frac{i}{M}\,\psi'_{1\,n,\ell}-i\,\omega\,\rho\,\psi_{1\,n,\ell}.\label{osci:20}\\
&&\psi_{4\,n,\ell}(\rho)=-\frac{|\ell|}{|\sigma|\,\sin\left(\sqrt{2\Lambda}\,\rho\right)} \,\psi_{1\,n,\ell}.\label{osci:21} \\
&&\psi_{5\,n,\ell}(\rho)=-\frac{k}{M}\,\psi_{1\,n,\ell}.\label{osci:22}
\end{eqnarray}

Equations (\ref{osci:17}) and (\ref{osci:18}) describe the relativistic energy eigenvalue and the wave function of the spinor field of the DKP oscillator field within the Bonnor-Melvin-Lambda solution. It's evident that the cosmological constant $\Lambda$ and the topological parameter $|\sigma|$ associated with the space-time geometry modify the eigenvalue solution. Additionally, the energy levels are influenced by the quantum numbers $\{n, \ell\}$ and change with variations in the oscillator frequency $\omega$. To elucidate the influence of these parameters on the energy levels, we've generated Figures 3-4.

In Figure 3, we depict the energy eigenvalue (\ref{osci:17}) while varying the values of the cosmological constant $\Lambda$ in Fig. 3(a), the topology parameter $|\sigma|$ in Fig. 3(b), the orbital quantum number $\ell$ in Fig. 3(c), and the frequency of oscillation $\omega$ in Fig. 3(d). It's evident that the energy level gradually increases in a linear manner. This increasing trend shifts downward with increasing values of $\Lambda$ and $|\sigma|$ in Figs. 3(a)-3(b). However, in Figs. 3(c)-3(d), this trend shifts upward with increasing values of $\ell$ and $\omega$.

Furthermore, in Figure 4, we present the energy spectrum concerning the cosmological constant $\Lambda$ in Fig. 4(a) and the topology parameter $|\sigma|$ in Fig. 4(b). Here, we observe a gradual decrease in the energy level with increasing values of $\Lambda$ and $|\sigma|$, respectively. This trend shifts downward with increasing values of $n$.

\section{Conclusions}

The study of both relativistic and non-relativistic limits of quantum systems in curved space has been a focal point in theoretical physics since long time. The gravitational field generated by curved space exerts significant influences on the quantum behaviors of spin-0 Bosonic fields, fermionic fields, and spin-1 fields. In addition, if the investigation is carried out in the presence of topological defects, then the eigenvalue solution of the quantum systems get modifications by these and break the degeneracy of the energy spectrum. Moreover, numerous authors introduced external uniform and non-uniform magnetic fields and presented the modified eigenvalue solution of these particles. In addition, many authors considered interaction potential of various kinds (both electromagnetic and non-electromagnetic potential) and showed that the energy eigenvalues of Bosonic and Fermionic fields get shifted. The eigenvalue solutions of the quantum system, obtained by solving the wave equations, undergo modifications due to the effects of the gravitational field, topological defects, interacting potential and external magnetic field and deviating the results from those obtained in flat space background. In this analysis, we focused on a four-dimensional magnetic space-time featuring a cosmological constant. There are only a handful Einstein-Maxwell solutions without and with cosmological constants have been known in the general relativity. The considered space-time is an example of electrovacuum Einstein-Maxwell solution in general relativity whose magnetic field strength is associated with two parameters $(|\sigma|, \Lambda)$ of the geometry. 

In {\it section 2}, we discussed the motion of spin-0 quantum particles described by the DKP equation and solved it in the background of magnetic space-time. We showed that the derived equation is the Bessel second-order differential equation whose solution is well-known. By imposing hard-wall confining potential condition, we determined the energy eigenvalue of spin-0 DKP fields given in equation (\ref{eq:29}). In {\it section 3}, we discussed the DKP-oscillator in the same magnetic space-time background. This oscillator field is studied  by a nonminimal substitution in the momentum operator. We derived the radial equation and solved it through the parametric Nikiforov-Uvarov method. We have seen that the energy eigenvalue of DKP oscillator fields given in equation (\ref{osci:17}) is influenced by the parameters $(|\sigma|, \Lambda)$ and the oscillator frequency $\omega$. We represented these energy profiles by illustrating a few graphs and have shown the nature of these by choosing different values of parameters $(|\sigma|, \Lambda)$ and the quantum number $n$.

In this context, intriguing results have been presented regarding the relativistic quantum system within the framework of a magnetic universe. The system under analysis exhibits the characteristics of a spin-0 Bosonic field and a Bosonic oscillator field, both influenced by a positive cosmological constant ($\Lambda>0$), indicative of a de-Sitter (dS) space background. The current phase of the universe's expansion is marked by acceleration, a phenomenon validated by the $\Lambda$CDM model. The interrelation between these theories remains an enigma in theoretical physics, presenting a compelling avenue for further exploration and understanding. 

Moving forward, future studies could focus into the interplay between this magnetic space-time and other fundamental forces, such as the strong and weak nuclear forces, within the context of quantum field theory and general relativity. Investigating the implications of magnetic space-time on phenomena like particle interactions, black hole dynamics, and cosmological evolution could provide invaluable insights into the nature of our universe and its fundamental constituents. Additionally, exploring the potential experimental signatures of magnetic space-time effects in high-energy particle physics experiments and astrophysical observations could offer empirical validation of theoretical predictions and pave the way for new discoveries in the field of theoretical physics.

\section*{Data Availability Statement }

No new data were analysed.

\section*{Conflict of Interest}

No conflict of interests in this paper.

\section*{Funding Statement }

No fund has received for this work.


\begin{thebibliography}{1}

\bibitem{k1} A. Einstein, Ann. Phys. (Berlin) {\bf{354}}, 769 (1916) \url{https://doi.org/10.1002/andp.19163540702}.

\bibitem{k2} B. P. Abbott, {\it et al.}, Phys. Rev. Lett. {\bf{116}}, 061102 (2016) \url{https://doi.org/10.1103/PhysRevLett.116.061102}.

\bibitem{k3} K. Akiyama, {\it et al.}, Astrophys. J. Lett. {\bf{875}}, L1 (2019) \url{https://doi.org/10.3847/2041-8213/ab0f43}.

\bibitem{k4} R. P. Feynman, and A. R. Hibbs,  {\it Quantum mechanics and path integrals}, Dover Publications Inc. (1965).

\bibitem{key-1} K. Gödel, Rev. Mod. Phys. \textbf{21}, 447 (1949) \url{https://doi.org/10.1103/RevModPhys.21.447}.

\bibitem{key-2} N. Drukker, B. Fiol and J. Simon, JCAP 10 ({\bf 2004}) 12 \url{https://doi.org/10.1088/1475-7516/2004/10/012}.

\bibitem{key-3} M. M. Som, and A. K. Raychaudhuri, Proc. R. Soc. A \textbf{304}, 81 (1968) \url{https://doi.org/10.1098/rspa.1968.0073}.

\bibitem{key-4} C. C. Barros Jr., Eur. Phys. J. C \textbf{42}, 119 (2005) \url{https://doi.org/10.1140/epjc/s2005-02252-7}.

\bibitem{key-5} F. Ahmed, Commun. Theor. Phys. \textbf{68}, 735 (2017) \url{ https://doi.org/10.1088/0253-6102/68/6/735}.

\bibitem{key-6} L. C. N. Santos, C. E. Mota, and C. C. Barros, Adv. High Energy Phys. \textbf{2019}, 2729352 (2019) \url{https://doi.org/10.1155/2019/2729352}.

\bibitem{key-7} G. de A. Marques, and V. B. Bezerra, Phys. Rev. D \textbf{66}, 105011 (2002) \url{https://doi.org/10.1103/PhysRevD.66.105011}.

\bibitem{key-8} A. Boumali, and N. Messai, Can. J. Phys. \textbf{92}, 1460 (2014) \url{https://doi.org/10.1134/S0202289321030026}.

\bibitem{key-9} E. R. Figueiredo Medeiros, and E. R. B. de Mello, Eur. Phys. J. C \textbf{72}, 2051 (2012) \url{https://doi.org/10.1140/epjc/s10052-012-2051-9}.

\bibitem{key-10} C. O. Edet et al., Results in Physics \textbf{39}, 105749 (2022) \url{https://doi.org/10.1016/j.rinp.2022.105749}.

\bibitem{key-11} F. Ahmed, Mol. Phys. {\bf 120}, e2124935 (2022) \url{https://doi.org/10.1080/00268976.2022.2124935}.

\bibitem{key-12} F. Ahmed, Phys. Scr. \textbf{98}, 015403 (2023) \url{https://doi.org/10.1088/1402-4896/aca6b3}.

\bibitem{key-13} R. J. Duffin, Phys. Rev. \textbf{54}, 1114 (1938) \url{https://doi.org/10.1103/PhysRev.54.1114}.

\bibitem{key-14} N. Kemmer, Proc. R. Soc. London, Ser. A \textbf{173}, 91 (1939) \url{https://doi.org/10.1098/rspa.1939.0131}.

\bibitem{key-15} G. Petiau, Acad. R. Belg. Cl. Sci. Mém. Collect. \textbf{8}, 16 (1939).

\bibitem{key-16} R. F. Guertin, Phys. Rev. D \textbf{15}, 1518 (1977) \url{https://doi.org/10.1103/PhysRevD.15.1518}.

\bibitem{key-17} Y. Nedjadi, and R. C. Barrett, J. Phys. G \textbf{19}, 87 (1993) \url{https://doi.org/10.1088/0954-3899/19/1/006}.

\bibitem{key-18} W. Greiner,  {\it Relativistic Quantum Mechanics: Wave Equations}, Springer, Berlin (2000).



\bibitem{kk1} M. Hosseinpour, and H. Hassanabadi. Adv. High Energy Phys. {\bf 2018}, 2959354 (2018) \url{https://doi.org/10.1140/epjc/s10052-018-5574-x}.

\bibitem{kk2} A. Boumali, A. Bouzenada, S. Zare, and H. Hassanabadi, Phys. A: Stat. Mech., \textbf{628}, 129134 (2023) \url{https://doi.org/10.1016/j.physa.2023.129134}.

\bibitem{kk3} S. Zare, H. Hassanabadi, and M. de Montigny. Gen. Relativ. Gravit., \textbf{52}, 25 (2020) \url{https://doi.org/10.1007/s10714-020-02676-0}.

\bibitem{kk4} S. Zare, H. Hassanabadi, and M. de Montigny, 	Int. J. Mod. Phys. A \textbf{35}, 2050195 (2020) \url{https://doi.org/10.1142/S0217751X2050195X}.

\bibitem{kk5} F. Ahmed, and H. Hassanabadi, Mod. Phys. Lett. A \textbf{35},2050031 (2020) \url{https://doi.org/10.1142/S0217732320500315}.

\bibitem{kk6} H. Sobhani, H. Hassanabadi, and W. S. Chung, Few-Body Syst.\textbf{61}, 7 (2020) \url{https://doi.org/10.1007/s00601-019-1537-5}.

\bibitem{MH5} M. de Montigny, M. Hosseinpour, and H. Hassanabadi, Int. J Mod. Phys. {\bf A 31}, 1650191 (2016) \url{https://doi.org/10.1142/S0217751X16501918}.

\bibitem{LBC} L. B. Castro, Eur Phys J C {\bf 75}, 287 (2015) \url{https://doi.org/10.1140/epjc/s10052-015-3507-5}.

\bibitem{MH2} L. B. Castro, Eur Phys J C {\bf 76}, 61 (2016) \url{https://doi.org/10.1140/epjc/s10052-016-3904-4}.

\bibitem{MH3} H. Hassanabadi, M. Hosseinpour, and M. de Montigny, Eur Phys J Plus {\bf 132}, 541 (2017) \url{https://doi.org/10.1140/epjp/i2017-11831-y}.

\bibitem{key-19} A. Boumali, and H. Aounallah, Adv. High Energy Phys. \textbf{2018}, 1031713 (2018) \url{https://doi.org/10.1155/2018/1031763}.

\bibitem{PRC} H. Hassanabadi, B. H. Yazarloo, S. Zarrinkamar, and A. A. Rajabi, Phys. Rev. C {\bf 84}, 064003 (2011) \url{https://doi.org/10.1103/PhysRevC.84.064003}.

\bibitem{CPC} S. Zarrinkamar, A. A. Rajabi, B. H. Yazarloo and H. Hassanabadi, Chin. Phys. C {\bf 37}, 023101 (2013) \url{https://doi.org/10.1088/1674-1137/37/2/023101}.

\bibitem{EPJP} S. Zarrinkamar, S. F. Forouhandeh, B. H. Yazarloo and H. Hassanabadi, Eur. Phys. J. Plus {\bf 128}, 109 (2013) \url{https://doi.org/10.1140/epjp/i2013-13109-x}.

\bibitem{IJTP} Y. Kasri, and L. Chetouani, Int J Theor Phys {\bf 47}, 2249 (2008) \url{https://doi.org/10.1007/s10773-008-9657-6}. 

\bibitem{CJP}  H. Hassanabadi, S. F. Forouhandeh, H. Rahimov, S. Zarrinkamar, and B. H. Yazarloo, Can. J Phys. {\bf 90}, 299 (2012) \url{https://doi.org/10.1139/p2012-019}.

\bibitem{FBS} H. Hassanabadi, M. Hosseini, S. Zare, and M. Hosseinpour, Few-Body Syst {\bf 60}, 12 (2019) \url{https://doi.org/10.1007/s00601-018-1480-x}.

\bibitem{EPL2} M. Darroodi, H. Mehraban, and S. Hassanabadi, EPL {\bf 118}, 10002 (2017) \url{https://doi.org/10.1209/0295-5075/118/10002}.

\bibitem{cor1}  V. Ya. Fainberg and B. M. Pimentel, Braz. J. Phys. {\bf 30}, 275 (2000) \url{https://doi.org/10.1590/S0103-97332000000200008}.

\bibitem{cor2} R. A. Krajcik and M. M. Nieto, Am. J. Phys. \textbf{45}, 818 (1977) \url{https://doi.org/10.1119/1.11054}.

\bibitem{k18} T. Gutsunaev, and V. Manko, Phys. Lett. {\bf A 123}, 215 (1987) \url{https://doi.org/10.1016/0375-9601(87)90063-6}.

\bibitem{k19} T. Gutsunaev, and V. Manko, Phys. Lett. {\bf A 132}, 85 (1988) \url{https://doi.org/10.1016/0375-9601(88)90257-5}.

\bibitem{k20} W. B. Bonnor, Proc. Phys. Soc. {\bf A 67}, 225 (1954) \url{https://doi.org/10.1088/0370-1298/67/3/305}.

\bibitem{k21} M. Melvin, Phys. Lett. {\bf 8}, 65 (1964) \url{https://doi.org/10.1016/0031-9163(64)90801-7}.

\bibitem{k22} M. Žofka, Phys. Rev. {\bf D 99}, 044058 (2019) \url{https://doi.org/10.1103/PhysRevD.99.044058}.

\bibitem{k23} L. Parker, Phys. Rev. Lett. {\bf 44}, 1559 (1980) \url{https://doi.org/10.1103/PhysRevLett.44.1559}. 

\bibitem{k25} E. Elizalde, Phys. Rev. {\bf D 36}, 1269 (1987) \url{https://doi.org/10.1103/physrevd.36.1269}.

\bibitem{k26} S. Chandrasekhar, Proc. R. Soc. Lond. A {\bf 349}, 571 (1976) \url{https://doi.org/10.1098/rspa.1976.0090}.

\bibitem{k27} L. C. N. Santos, and C. C. Barros Jr., Eur. Phys. J. C {\bf 77}, 186 (2017) \url{https://doi.org/10.1140/epjc/s10052-017-4732-x}.

\bibitem{k28} L. C. N. Santos, and C. C. Barros Jr., Eur. Phys. J. C {\bf 78}, 13 (2018) \url{https://doi.org/10.1140/epjc/s10052-017-5476-3}.

\bibitem{k29} R. L. L. Vitória, and K. Bakke, Eur. Phys. J. C {\bf 78}, 175 (2018) \url{https://doi.org/10.1140/epjc/s10052-018-5658-7}.

\bibitem{k30} F. Ahmed, Int. J. Mod. Phys. {\bf A 37}, 2250186 (2022) \url{https://doi.org/10.1142/S0217751X2250186X}.

\bibitem{k31} F. Ahmed, Commun. Theor. Phys.{\bf 75}, 025202 (2023) \url{https://doi.org/10.1088/1572-9494/aca650}.

\bibitem{k35} A. Bouzenada, A. Boumali, Ann. Phys. (NY) \textbf{452}, 169302 (2023) \url{https://doi.org/10.1016/j.aop.2023.169302}.

\bibitem{k35-1} A. Bouzenada, A. Boumali, R. L. L. Vitoria, F. Ahmed, and M. Al-Raeei, Nucl. Phys. B \textbf{994}, 116288 (2023) \url{https://doi.org/10.1016/j.nuclphysb.2023.116288}.

\bibitem{k35-2} A. Bouzenada, A. Boumali, and F. Serdouk, Theor. Math. Phys,\textbf{216}, 1055 (2023) \url{https://doi.org/10.1134/S0040577923070115}.

\bibitem{k35-3} A. Bouzenada, A. Boumali, and E. O. Silva, Ann. Phys. (N. Y.) \textbf{458}, 169479 (2023) \url{https://doi.org/10.1016/j.aop.2023.169479}.

\bibitem{k36} V. B. Bezerra, M. S. Cunha, L. F. F. Freitas, C. R. Muniz and M. O. Tahim, Mod. Phys. Lett. {\bf A 32}, 1750005 (2016) \url{https://doi.org/10.1142/S0217732317500055}.

\bibitem{k37} L. C. N. Santos, and C. C. Barros Jr., Int. J. Mod. Phys. {\bf A 33}, 1850122 (2018) \url{https://doi.org/10.1142/S0217751X18501221}.

\bibitem{k38} E. O. Pinho, and C. C. Barros Jr., Eur. Phys. J. C. {\bf 83}, 745 (2023) \url{https://doi.org/10.1140/epjc/s10052-023-11907-y}.

\bibitem{k39} P. Sedaghatnia, H. Hassanabadi, and F. Ahmed, Eur. Phys. J. C. {\bf 79}, 541 (2019) \url{https://doi.org/10.1140/epjc/s10052-019-7051-6}.

\bibitem{k40} A. Guvendi, S. Zare, and H. Hassanabadi, Phys. Dark Univ. {\bf 38}, 101133 (2022) \url{https://doi.org/10.1016/j.dark.2022.101133}. 

\bibitem{k41} R. L. L. Vitória, and K. Bakke, Eur. Phys. J. Plus {\bf 133}, 490 (2018) \url{https://doi.org/10.1140/epjp/i2018-12310-9}.

\bibitem{k42} L. C. N. Santos, and C. C. Barros Jr., Eur. Phys. J. C {\bf 76}, 560 (2016) \url{https://doi.org/10.1140/epjc/s10052-016-4409-x}.

\bibitem{MA2} M. Astorino, JHEP 06 ({\bf 2012}) 086 \url{https://doi.org/10.1007/JHEP06(2012)086}.





\bibitem{MA} M. Abramowitz, and I. A. Stegun, {\tt Handbook of Mathematical Functions with Formulas, Graphs, and Mathematical Tables}, New York: Dover (1972).

\bibitem{GBA} G. B. Arfken, H. J. Weber and F. E. Harris, {\it Mathematical Methods for Physicists}, Elsevier (2012).

\bibitem{KB} K. Bakke, Cent. Eur. J. Phys. {\bf 11}, 1589 (2013) \url{https://doi.org/10.2478/s11534-013-0313-2}.

\bibitem{KB2} K. Bakke, Mod. Phys. Lett. {\bf B 27}, 1350018 (2013) \url{https://doi.org/10.1142/S0217984913500188}.

\bibitem{KB3} K. Bakke, Ann. Phys. (NY) {\bf 346}, 51 (2014) \url{https://doi.org/10.1016/j.aop.2014.04.003}.

\bibitem{KB5} A. V. D. M. Maia, and K. Bakke, Phys. B {\bf 531}, 213 (2018) \url{https://doi.org/10.1016/j.physb.2017.12.045}.

\bibitem{KB4} K. Bakke, Int. J. Theor. Phys. {\bf 54}, 2119 (2015) \url{https://doi.org/10.1007/s10773-014-2418-9}.

\bibitem{KB6} K. Bakke, Eur. Phys. J. B {\bf 85}, 354 (2012) \url{https://doi.org/10.1140/epjb/e2012-30490-6}.

\bibitem{KB7}  K. Bakke, C. Furtado, Eur. Phys. J. B {\bf 87}, 222 (2014) \url{https://doi.org/10.1140/epjb/e2014-50106-5}.

\bibitem{KB8} K. Bakke, H. Belich, J. Phys. G: Nucl. Part. Phys. {\bf 42}, 095001 (2015) \url{https://doi.org/10.1088/0954-3899/42/9/095001}.

\bibitem{RLLV5} E. A. F. Bragança, R. L. L. Vitória, H. Belich, and E. R. Bezerra de Mello, Eur. Phys. J. C {\bf 80}, 206 (2020) \url{https://doi.org/10.1140/epjc/s10052-020-7774-4}.

\bibitem{SR} F. Ahmed, Sci. Rep. {\bf 12}, 8794 (2023) \url{https://doi.org/10.1038/s41598-022-12745-w}.
 
\bibitem{CTP} F. Ahmed, Commun. Theor. Phys. 72 (02), 025103 (2020) \url{https://doi.org/10.1088/1572-9494/ab6187}.

\bibitem{MH} M. Hosseinpour, H. Hassanabadi, and F. M. Andrade, Eur Phys J C {\bf 78}, 93 (2018) \url{https://doi.org/10.1140/epjc/s10052-018-5574-x}.

\bibitem{AB2} A. Boumali, J. Math. Phys. {\bf 49}, 022302 (2008) \url{https://doi.org/10.1063/1.2841324}, Erratum: {\it ibid} {\bf 54}, 099902 (2013) \url{https://doi.org/10.1063/1.4821200}.

\bibitem{NC} N. Candemir, and F. Ahmed, Phys. Scr. {\bf 98}, 065224 (2023) \url{https://doi.org/10.1088/1402-4896/acd669}.

\bibitem{JMP2} M. Falek, and M. Merad, J. Math. Phys. {\bf 51}, 033516 (2010) \url{https://doi.org/10.1063/1.3326236}.

\bibitem{JMP3} M. Falek, and M. Merad, J. Math. Phys. {\bf 50}, 023508 (2009) \url{https://doi.org/10.1063/1.3076900}.

\bibitem{EPJC} H. Hassanabadi, Z. Molaee, and S. Zarrinkamar, Eur. Phys. J. C {\bf 72}, 2217 (2012) \url{https://doi.org/10.1140/epjc/s10052-012-2217-5}. 

\bibitem{GDM} G. R. de Melo, M. de Montigny, and E. S. Santos, J. Phys.: Conf. Ser. {\bf 343}, 012028 (2012) \url{https://doi.org/10.1088/1742-6596/343/1/012028}.

\bibitem{MF} M. Falek, and M. Merad, Commun. Theor. Phys. {\bf 50}, 587 (2008) \url{https://doi.org/10.1088/0253-6102/50/3/10}.

\bibitem{AFN} A. F. Nikiforov, and V. B. Uvarov, {\it Special Functions of Mathematical Physics}, Birkhauser (1988) \url{https://doi.org/10.1007/978-1-4757-1595-8}.

\bibitem{PHYSSR} A. Boumali, Phys. Scr. {\bf 76}, 669 (2007) \url{https://doi.org/10.1088/0031-8949/76/6/014}.

\bibitem{PLA} L. B. Castro, and A. S. de Castro, Phys. Lett. A {\bf 375}, 2596 (2011) \url{https://doi.org/10.1016/j.physleta.2011.05.067}.


\end{thebibliography}
\end{document}